\documentclass[cernpreprint, atlasdraft=false, USenglish, texmf, orcidlogo]{atlasdoc}
 
\usepackage{atlaspackage}
\usepackage{atlasbiblatex}
 
\usepackage{atlasphysics}

\graphicspath{{logos/}{figures/}}
 
\usepackage{ANA-SUSY-2020-04-PAPER-defs}
 
\usepackage{adjustbox}

\AtlasTitle{Search for Nearly Mass-Degenerate Higgsinos Using Low-Momentum Mildly Displaced Tracks in $pp$ Collisions at $\sqrt{s}=13$~TeV with the ATLAS Detector}

\AtlasAbstract{
Higgsinos with masses near the electroweak scale can solve the hierarchy problem and provide a dark matter candidate, while detecting them at the LHC remains challenging if their mass splitting is $\mathcal{O}(1~\text{GeV})$.
This Letter presents a novel search for nearly mass-degenerate Higgsinos in events with an energetic jet, missing transverse momentum, and a low-momentum track with a significant
transverse impact parameter using 140~fb$^{-1}$ of proton-proton collision data at $\sqrt{s}=13$~TeV collected by the ATLAS experiment.
For the first time since LEP, a range of mass splittings between the lightest charged and neutral Higgsinos from $0.3$~GeV to $0.9$~GeV is excluded at 95\% confidence level, with a maximum reach of approximately $170$~GeV in the
Higgsino mass.
}

\author{The ATLAS Collaboration}
 
\AtlasRefCode{SUSY-2020-04}
 
\PreprintIdNumber{CERN-EP-2024-012}
 
\AtlasDate{\today}

\AtlasJournalRef{Phys. Rev. Lett. 132 (2024) 221801}
\AtlasDOI{10.1103/PhysRevLett.132.221801}

\hypersetup{pdftitle={ATLAS document},pdfauthor={The ATLAS Collaboration}}
 
\begin{document}
 
\maketitle

The \textit{natural} solution from supersymmetry~\cite{myGolfand:1971iw,Volkov:1973ix,Wess:1974tw,Wess:1974jb,Ferrara:1974pu,Salam:1974ig} to the hierarchy problem~\cite{Barbieri:1987fn,deCarlos:1993yy} provides a strong motivation to search for light Higgsinos, the fermionic partners of the Higgs bosons.
Since the Higgsino mass is connected to the electroweak symmetry breaking scale in the Minimal Supersymmetric Standard Model (MSSM)~\cite{Fayet:1976et,Fayet:1977yc}, it is generally favored to be near the electroweak scale ($\mathcal{O}(100~\GeV)$) in this framework.
This light Higgsino scenario is also appealing because it provides a viable Dark Matter (DM) candidate if $R$ parity~\cite{Farrar:1978xj} is conserved, and a neutral Higgsino is the stable lightest supersymmetric particle with a mass below 1.1~\TeV~\cite{CIRELLI2007152,Kowalska_2018}.
These considerations indicate that Higgsinos may be within the mass reach of the Large Hadron Collider (LHC), and the direct production of Higgsinos has been a key probe for the search program.

The Higgsino phenomenology highly depends on their mass hierarchy with respect to the wino and bino, the fermionic partners of the $SU(2)_L$ and $U(1)_Y$ gauge boson in the SM.
In the MSSM, the Higgsino ($\widetilde{H}$) and wino ($\widetilde{W}$), and bino ($\widetilde{B}$) fields, whose masses are characterized by the parameters $\mu$, $M_1$, and $M_2$, respectively, undergo kinetic mixing as the result of the electroweak symmetry breaking,
yielding charged ($\Cone, \Ctwo$) and neutral ($\None, \Ntwo, \Nthree, \Nfour$) mass eigenstates, where the order of the subscripts indicates increasing mass.
In the case where $|\mu| \ll |M_1|,|M_2|$, the lightest charged and two neutral Higgsino-like eigenstates form a nearly mass-degenerate triplet of Higgsino-like mass eigenstates (\Cone, \Ntwo, \None), collectively referred to as Higgsinos in this Letter.
In the pure Higgsino limit where the bino and wino are decoupled in mass, $\min{(|M_1|, |M_2|)} > \mathcal{O}(10~\TeV)$, radiative corrections induce a small mass splitting ($\dmpm \approx 250\text{--}400~\MeV$~\cite{nagata2015higgsino}) between the Higgsinos that gives the \Cone a long enough lifetime to produce the disappearing track signature in the detector, as exploited by existing searches~\cite{SUSY-2016-06,SUSY-2018-19,CMS-EXO-16-044,CMS-EXO-19-010}.
On the other hand, mass splittings larger than a few GeV are possible when either the bino or wino states are light, $\min{(|M_1|, |M_2|)} < \mathcal{O}(1~\TeV)$, enabling low-momentum prompt lepton searches to target the $\Ntwo \ra Z^* \None \ra \ell^+ \ell^- \None$ decay and provide constraints down to $\dmn \approx 2~\GeV$~\cite{SUSY-2018-16,SUSY-2019-09,CMS-SUS-19-012}.
However, a distinct sensitivity gap remains for $\dmpm \approx 0.3\text{--}1~\GeV$, where the strongest constraints are still set by the LEP experiments~\cite{LEPlimits}.
Importantly, this mass-splitting range is challenging for direct DM detection experiments due to the vanishing DM-nuclei coupling~\cite{Fukuda:2020}, while the $\dmn > 2~\GeV$ and pure Higgsino scenarios are highly constrained by those searches~\cite{nagata2015higgsino}.
This Letter reports a new search by the ATLAS\footnote{\AtlasCoordFootnote} experiment~\cite{PERF-2007-01} to address this sensitivity gap.
Following the proposal in Ref.~\cite{Fukuda:2020} and described in detail below, the search identifies low-momentum charged particles that are consistent with the decay products of a $\Cone$ that has a discernible flight length from the proton-proton ($pp$) collision point.
 
For the mass splitting of interest of $\dmpm \approx 0.3\text{--}1~\GeV$, the decay flight length of $\Cone$ can reach $\mathcal{O}(0.1\text{--}1~\text{mm})$. Charged decay particles from $\Cone$, dominated by charged pions, have a transverse momentum (\pt) of a few GeV, offering a resonable reconstruction efficiency.
The transverse impact parameter, $\dzero$, with its resolution, $\sigma(\dzero)$, is the distance of closest approach of the charged particle trajectory in the transverse ($x$\nobreakdash--$y$) plane with respect to the centroid of the $pp$ collision of interest. Since the displacement of the $\Cone$ decay position from the $pp$ collision point is correlated to $\dzero$, the transverse impact parameter significance, $\sdzero = |\dzero|/\sigma(\dzero)$, can be the the most distinctive discriminant to identify the decay of $\Cone$ together with an adequate requirement of \pt.
The $\Cone$ flight length is sufficiently short so that most of the decay charged particles are expected to pass through the innermost tracking layer. Satisfying such a decay condition with sufficiently large $\sdzero$ is referred to as ``mildly displaced'' in this Letter.
The production of a Higgsino-pair system with initial-state radiation (ISR) can be used to trigger the event based on missing transverse momentum (denoted as \ptmissvec along with its magnitude \met).
Moreover, the additional Lorentz boost by the ISR recoil enhances both \pt and $\sdzero$, so that more decay charged particles get mildly displaced, which increases the sensitivity to even smaller mass-splitting values.
While the conventional ``monojet'' searches~\cite{EXOT-2018-06,CMS-EXO-20-004} that probe this ISR event topology for generic DM production at the LHC do not
provide significant sensitivity to Higgsino production due to the overwhelming Standard Model (SM) background,
the inclusion of a displaced track requirement allows for a significant reduction of these backgrounds, which allows the exploration of this range of mass-splitting values for the first time since LEP.
This search uses the $pp$ collision data collected at the LHC during 2015--2018 at a center-of-mass energy of $\sqrt{s}=13~\TeV$, corresponding to an integrated luminosity of 140~\ifb.
 
An example signal diagram of the target signature is shown in Figure~\ref{fig:diagram}.
The same $R$-parity conserving Higgsino simplified model is considered as in Ref.~\cite{SUSY-2018-16,SUSY-2019-09} where the mass of the \Cone is halfway between that of the \Ntwo and the \None, i.e.\ $m(\Ntwo)-m(\Cone) = m(\Cone)-m(\None)$.
The Higgsino-pair production modes considered are \ConepConem, \ConeNone, \ConeNtwo, and \NtwoNone.
The largest branching ratio of \Cone (\Ntwo) decays is to a single $\pi^\pm$ ($\pi^0$) when $\dmpm = 0.3\text{--}1~\GeV$~\cite{pasechnik2009neutralino}; about 80\% of \Cone decays to $\pi^\pm\None$ with $\dmpm=0.5~\GeV$; and about 70\% of \Ntwo decays to $\pi^0\None$ with $\dmn=0.5~\GeV$.
The identified displaced track in signal events therefore typically corresponds to a $\pi^\pm$ from a \Cone decay,
but a small fraction can also arise from $\Cone \ra e\nu\None$, $\Cone \ra \mu\nu\None$, $\Ntwo \ra e^+e^-\None$, $\Ntwo \ra \mu^+\mu^-\None$ and $\Ntwo \ra \pi^+\pi^-\None$ decays; all are taken into account as signal in the analysis.
 
\begin{figure}[t]
\centering
\includegraphics[width=0.32\textwidth, keepaspectratio]{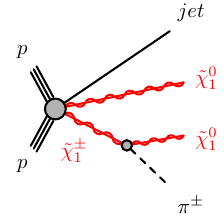}
\caption{Example signal diagram for the targeted signature featuring a jet from initial-state radiation. For illustration, the \ConeNone process is shown, while the production of \ConepConem, \ConeNtwo, and \NtwoNone is considered in the search as well.
}
\label{fig:diagram}
\end{figure}

The ATLAS experiment is a multipurpose particle detector with a forward-backward symmetric cylindrical geometry and nearly $4\pi$ coverage in solid angle.
It consists of an inner tracking detector (ID) surrounded by a superconducting solenoid, sampling electromagnetic (EM) and hadronic calorimeters, and a muon spectrometer (MS) with three toroidal superconducting magnets.
Charged particle tracks are reconstructed using the hits in the ID and are required to have $\pt > 500~\MeV$.
For tracks with $\pt=2~\GeV$, the intrinsic resolution on $d_0$ is approximately 0.05~mm, which improves to 0.03~mm at $\pt=5~\GeV$ and 0.01~mm at $\pt > 10~\GeV$~\cite{ATL-IDTR-2018-008}.
A two-level trigger system is used to select events for storage.
The events in the main dataset used in this analysis relied on the \met trigger~\cite{TRIG-2019-01},
while the auxiliary dataset for the background estimation and validation was collected using the single-electron~\cite{TRIG-2018-05} or single-photon triggers~\cite{TRIG-2018-05}.
An extensive software suite~\cite{ATL-SOFT-PUB-2021-001} is used in data simulation, in the reconstruction
and analysis of real and simulated data, in detector operations, and in the trigger and data acquisition
systems of the experiment.
 
The $pp$ interaction vertex with the highest $\pt^2$ sum of
associated tracks is selected as the hard-scatter vertex of interest.
Hadronic jets are reconstructed from particle-flow objects~\cite{PERF-2015-09} calibrated at the EM scale using the anti-$k_t$ algorithm~\cite{Cacciari:2008gp,Fastjet} with a radius parameter of $R=0.4$.
Jets with $\pt > 20~\GeV$ and $|\eta|<2.8$ are considered in the analysis.
To suppress the contribution from additional $pp$ collisions in the same and neighboring bunch crossings (``pileup''), jets with $\pt < 60~\GeV$ and $|\eta| < 2.5$ are required to pass the ``tight'' working point of the jet vertex tagger~\cite{PERF-2014-03}.
 
Electrons, muons, and photons are reconstructed and selected mainly to define the events used for the background estimation and validation.
Electrons are reconstructed from energy deposits in the EM calorimeter associated with tracks found in the ID.
Muons are reconstructed by combining tracks reconstructed in the ID with tracks or track segments found in the MS.
Electrons (muons) must have $\pt > 4.5~(3)~\GeV$, be reconstructed within $|\eta|<2.47~(2.7)$, and satisfy the tight~\cite{EGAM-2018-01} (medium~\cite{MUON-2018-03}) identification criterion.
To ensure that selected electrons and muons (collectively referred to as leptons in this Letter and denoted by $\ell$) originate from the selected hard-scatter vertex,
their tracks are required to fulfill $|z_0 \sin{\theta}|<0.5~\text{mm}$ where $z_0$ is the longitudinal impact parameter.
Photons are reconstructed as EM clusters with either no matching ID track or with a matching conversion vertex from $e^{+}e^{-}$ pairs in the ID material~\cite{EGAM-2018-01}.
Photon candidates with $\pt>25~\GeV$ and $|\eta|<2.37$ passing the tight identification criterion~\cite{EGAM-2018-01} are selected.
 
To prevent the reconstruction of a single particle as multiple objects,
an overlap removal procedure described in Ref.~\cite{SUSY-2018-41}
is applied to the leptons, photons, and jets.
The \met is calculated as the magnitude of the negative vectorial sum of the transverse momenta of all leptons, photons and jets calibrated to their respective energy scales, and an additional soft term constructed from tracks originating from the hard-scatter vertex but not associated with any of the reconstructed objects~\cite{PERF-2016-07}.
Jets in the forward direction (up to $|\eta|<4.5$) are also considered in the \met calculation.
 
Finally, the low-\pt charged track from the \Cone or \Ntwo decays, referred to as the signal candidate track, is defined as an ID track associated with the hard-scatter vertex
that fulfills: $2~\GeV < \pt < 5~\GeV$, $|\eta|<1.5$, $|d_0|<10~\text{mm}$, $|z_0\sin{\theta}|<3~\text{mm}$, and the ``tight primary'' track selection criteria defined in Ref.~\cite{ATL-PHYS-PUB-2015-051}.
The loose requirement on the track $|d_0|$ retains efficiency for charged particles with long lifetimes, while the tight primary requirement ensures high quality tracks based on the number of hits in the silicon detectors of the ID system.
A signal candidate track must also have a measurement in the first layer of the ID at a radius of approximately $33~\text{mm}$, and not be associated with any secondary $K_S^0$ or $\Lambda$ decay vertices as identified
by an algorithm optimized to reconstruct two-body V0 decays~\cite{STDM-2010-09,IDTR-2021-03}.
An isolation requirement is applied such that no other tight primary tracks satisfying $\pt > 1~\GeV$, $|d_0|<10~\text{mm}$, and $|z_0\sin{\theta}|<3~\text{mm}$ are allowed within an angular distance of $\Delta R < 0.4$ with respect to the signal candidate track.
The signal candidate track needs to be aligned with the \ptmissvec by requiring $\dphimettrack < 0.4$, reflecting the targeted event topology in which the Higgsinos are collimated due to the transverse boost by the recoiling ISR jet.
If multiple tracks pass this selection, the track with the highest \sdzero is selected as the signal candidate track.
 
Monte Carlo~(MC) simulations are used to model the Higgsino signals and the SM backgrounds,
where the generation of all simulated event samples includes the effect of multiple $pp$ interactions per bunch crossing, as well as the effect on the detector response due to interactions from bunch crossings before or after the one containing the hard interaction.
Details about the event simulation configurations  used can be found in Appendix A.
For signal samples the lifetime and branching ratios of the \Cone and \Ntwo are calculated following the prescription described in Ref.~\cite{pasechnik2009neutralino} and propagated to the simulation of their decays during the event generation.
The signal cross sections are computed at next-to-leading order (NLO) in the strong coupling constant,
adding the resummation of soft-gluon emission at next-to-leading-logarithm accuracy~\cite{Beenakker:1999xh,Debove:2010kf,Fuks:2012qx,Fuks:2013vua,Fiaschi:2018hgm,Borschensky:2014cia}.
The \mbox{PDF4LHC15\_mc} parton distribution function (PDF) set is used following the recommendations in Ref.~\cite{Butterworth:2015oua}.
The production cross section is 3.95~pb for $m(\Ntwo,\Cone,\None)=(151,150.5,150)~\GeV$ with all of the production modes (\ConepConem, \ConeNonetwo, \NtwoNone) included.

The dominant SM background originates from events with a leptonically decaying $W(\ra \ell\nu/\tau\nu)$ or $Z(\ra \nu\nu/\tau\tau)$ boson associated with jets, referred to as $W+\text{jets}$ and $Z+\text{jets}$, respectively.
Other backgrounds processes include diboson,
\ttbar, and single top-quark production.
The production of QCD multijet, triboson, and other rare processes including top quarks were found to be negligible.
 
The signal region (SR) is defined by selecting events that have no leptons or photons after the \met trigger requirement.
A set of event cleaning criteria is applied~\cite{DAPR-2012-01}, including a veto of events with cosmic muons or muons with poor \pt determination.
The events must also contain at least one jet with $\pt>250~\GeV$ and $|\eta|<2.4$, but no more than four jets to suppress the \ttbar background.
To ensure the removal of beam-induced background, events are rejected if the highest \pt jet has an anomalous energy profile in the calorimeter that fails the tight cleaning criteria~\cite{ATLAS-CONF-2015-029}.
The azimuthal angular separation between \ptmissvec and all jets must satisfy $\dphimetjet > 0.4$ to suppress the QCD multijet background.
Additionally, events are required to have $\met>600~\GeV$ in order to reject the bulk of the SM backgrounds, primarily from $W/Z+\text{jets}$ production.
This offline \met requirement is well above $200~\GeV$, where the \met trigger is already found to be fully efficient.
Finally, the presence of a signal candidate track with a transverse displacement significance of $\sdzero>8$ is required.
The \sdzero selection efficiency is approximately 40~\% for tracks originating from Higgsino decays in a signal MC sample with $\dmpm = 0.5~\GeV$.
The SR is split into two bins labeled SR-Low ($8 < \sdzero < 20$) and SR-High ($\sdzero>20$) to maintain sensitivity to a range of mass splittings.
With all of the selection criteria applied, the acceptance times efficiency in the union of SR-Low and SR-High is roughly $2.1\times10^{-4}$ for the Higgsino signal model with $m(\None) = 150~\GeV$ and $\dmpm = 0.5~\GeV$.
 
The leading SM background is $W(\ra \tau\nu)+\text{jets}$ production where a pion or lepton from a low-\pt $\tau$-lepton decay is tagged as the signal candidate track.
This background is referred to as the ``$\tau$ track'' background. Such tracks tend to exhibit a harder \pt spectrum compared to the signals of interest, and their contribution to the SR is largely reduced by the track $\pt < 5~\GeV$ requirement.
To estimate this background a partially data-driven method is used where the MC sample is normalized to control regions (CRs) with higher track \pt ($8\text{--}20~\GeV$) with respect to the SR.
A CR with exactly zero (one) leptons is used to constrain the hadronic (leptonic) $\tau$ decay background yield.
Both $\tau$ track background MC samples share the same normalization factor.
A test was performed in which separate normalization factors were derived for the hadronic and leptonic $\tau$ decay backgrounds, resulting in a consistent prediction of their event yields in the SR bins compared to the use of a single normalization factor.

The subleading SM background is $W/Z+\text{jets}$ events where the signal candidate tracks arise from hadrons with measurably long lifetimes (e.g. $\Lambda$, $K_S^0$) in pileup jets or the underlying event, referred to as the ``QCD track'' background. These tracks tend to dominate at low \pt, motivating the $\pt > 2~\GeV$ requirement for signal tracks that significantly reduces their contribution to the SR.
A fully data-driven method is employed to estimate this background.
First, the shape of the \sdzero distribution is extracted from the data events in a control region with exactly one muon (CR-1$\mu$), which is dominated by $W(\ra \mu\nu)+\text{jets}$.
This sample is then normalized to the data in a low-\sdzero control region (CR-0$\ell$) to obtain the estimate in the SR, where CR-0$\ell$ is defined to be adjacent to the SR by inverting the \sdzero selection (i.e. $\sdzero<8$).
This method relies on the fact that the shape of the \sdzero distribution is dictated by the breakdown of the hadron components in the pileup jets or underlying events, which do not strongly depend on the details of the hard collision process.
Control region CR-1$\mu$ has the same selection as the SR except that exactly one muon is required, and that $\ptrecoil > 300~\GeV$ is applied instead of $\met>600$, where $\ptrecoil =|\ptmuvec + \ptmissvec|$ is the proxy for the \pt of the $W$ boson.
The loosened selection allows roughly a factor of 30 increase in the data yield in the CR.
Contributions from SM processes other than $W(\ra \mu \nu)+\text{jets}$ are subtracted when extracting the shape of the \sdzero distribution.
 
Backgrounds other than the $\tau$ track and the QCD track background are minor, and are estimated using the MC simulation.
 
The background modeling is tested by comparing the estimates and the data in dedicated validation regions (VRs).
The $\tau$ track background estimate is validated in regions with an intermediate track \pt ($5\text{--}8~\GeV$) adjacent to the SR ($2\text{--}5~\GeV$),
with a shifted \met requirement of $300~\GeV < \met < 400~\GeV$ to increase the data yield and to suppress signal contamination.
For the QCD track background estimate, the validity of the \sdzero shape as extracted from $W(\ra \mu \nu)+\text{jets}$ events is verified in dedicated VRs defined by requiring either one electron, two leptons, or one photon, which are dominated by $W(\ra e\nu)+\text{jets}$, $Z(\ra \ell\ell)+\text{jets}$, and $\gamma+\text{jets}$ production, respectively. Here, the signal candidate tracks have an origin similar to that of QCD track events in CR-1$\mu$ and the SR.
An additional VR populated with both the $\tau$ track and QCD track backgrounds is defined near the SR to test the estimate of both components (VR-0$\ell$-low\met).
This VR is defined by shifting the \met selection with respect to the SR to $300~\GeV<\met<400~\GeV$.
A more detailed description of the VRs is given in Appendix B, and schematic illustrations of the region layout can be found in the Supplemental Material~\cite{SupplementalMaterial} of this Letter.
 
To obtain the final background estimates, the extraction and the application of the \sdzero shape for the QCD track background and the normalization of the $\tau$ track background are performed via a simultaneous profile log-likelihood fit~\cite{Cowan:2010js_paperAndErr} including all CRs.
SRs and VRs are not included in the fit as constraints to estimate the backgrounds.
Systematic uncertainties are implemented in the fit as nuisance parameters constrained by Gaussian probability density functions.
Correlations between systematic uncertainties arising from common sources are maintained across processes and regions.
Instrumental and theoretical uncertainties are assigned on the modeling of signals and backgrounds except for the QCD track background.
The instrumental uncertainties include those on lepton trigger, reconstruction, and identification efficiencies~\cite{MUON-2018-03,PERF-2017-01};
lepton energy scale and resolution~\cite{MUON-2018-03,PERF-2017-03};
jet energy scale and resolution~\cite{JETM-2018-05};
jet vertex tagging~\cite{ATLAS-CONF-2014-018};
modeling of \met~\cite{PERF-2016-07} and pileup;
and integrated luminosity~\cite{Avoni:2018iuv}.
Theoretical uncertainties include the cross section uncertainties and the shape uncertainties
due to the renormalization and factorization scales~\cite{ATL-PHYS-PUB-2016-002}, PDFs~\cite{Butterworth:2015oua}, and parton showering. 
For the data-driven estimate on the QCD track background, uncertainties of 5\% and 10\% are assigned to SR-Low and SR-High, respectively,
to account for any potential \sdzero shape difference between CR-1$\mu$ and the SR and VRs due to the varying physics processes involved and kinematic selections, as evaluated in MC simulation.
The fit also includes the systematic uncertainties due to limited MC simulation sample sizes and the data sample size in the CRs, which are the dominant contributions among all systematic uncertainties in the SR and VRs.
The background estimate is found to agree well with the observed data in all of the VRs within the uncertainty.
 
The numbers of observed events in each SR bin are summarized in Table~\ref{tab:yields}, along with the SM background predictions.
The MC normalization factor for the $\tau$ track background is found to be $1.1 \pm 0.1$, derived from a fit under the background-only hypothesis.
The postfit \sdzero distribution in the SR is shown in Figure~\ref{fig:SRplots},
where no significant deviation from the expectation is observed.
The absence of a data excess is translated into exclusion limits at 95\% confidence level (CL) on the Higgsino simplified model using the \CLs prescription~\cite{Read:2002hq} employing the asymptotic formulae~\cite{Cowan:2010js_paperAndErr} for the profile likelihood ratio.
The two SR bins are included as constraints in the fit for computing exclusion limits. Including the SRs has a negligible effect on the fitted background yields.
The obtained expected and observed exclusion limits are shown in Figure~\ref{fig:limit}, where mass splittings in the range of $0.3~\GeV < \dmpm < 0.9~\GeV$ are excluded for a given value of $m(\Cone)$.
The search sensitivity peaks at $\dmpm = 0.6~\GeV$, for which $m(\Cone)$ is excluded up to approximately $170~\GeV$ in $m(\Cone)$.
The model-independent upper limits and constraints on generic physics processes beyond the Standard Model are also derived. These limits, assuming no signal contamination in the CRs, are presented in Table~\ref{tab:yields}.

% The next lines are included from the .//yield_table.tex input file
\begin{table}[t]
\caption{Number of expected and observed data events in the SR (top), and the model-independent upper limits obtained from their consistency (bottom).
The symbol $\tau_\ell$ ($\tau_h$) refers to fully leptonic (hadron-involved) $\tau$ decays.
The ``Others'' category includes contributions from minor background processes including $t\bar{t}$, single top and diboson.
The individual uncertainties can be correlated and do not necessarily sum up in quadrature to the total uncertainty.
The uncertainty associated with the expected number of events includes the systematic uncertainty and the statistical uncertainty due to the limit data sample size in the CRs.
The bottom section shows the observed 95\% CL upper limits on the visible cross section ($\langle\epsilon{\mathrm{\sigma}}\rangle_{\mathrm{obs}}^{95}$),
on the number of generic signal events ($S_{\mathrm{obs}}^{95}$) as well as the expected limit ($S_{\mathrm{exp}}^{95}$) given the expected number (and $\pm1\sigma$ deviations from the expectation) of background events.
}
\centering
\begin{tabular}{ll rr }
\toprule
&     & \multicolumn{1}{c}{SR-Low} & \multicolumn{1}{c}{SR-High} \\
\midrule
\multicolumn{2}{l}{Observed data} & $35$ & $15$ \\
\midrule
\multicolumn{2}{l}{SM prediction} & $37 \pm 4$ & $14.8 \pm 2.0$ \\
\cmidrule{2-4}
& QCD track                         & $14.0 \pm 1.7$ & $10.0 \pm 1.6$ \\
& $W(\ra \tau_\ell\nu)+\text{jets}$ & $9.6 \pm 1.6$  & $2.0 \pm 0.6$ \\
& $W(\ra \tau_h\nu)+\text{jets}$    & $10.6 \pm 2.0$ & $1.9 \pm 0.8$ \\
& Others                            & $3.2 \pm 0.7$  & $0.8 \pm 0.4$ \\
\bottomrule
\toprule
\multicolumn{2}{l}{$\langle\epsilon{\mathrm{\sigma}}\rangle_{\mathrm{obs}}^{95}$ [fb]} & $0.10$                       & $0.07$ \\
\multicolumn{2}{l}{$S_{\mathrm{obs}}^{95}$}                                            & $13.5$                       & $9.9$ \\
\multicolumn{2}{l}{$S_{\mathrm{exp}}^{95}$}                                            & ${ 15.1 }^{ +6.3 }_{ -4.2 }$ & $ { 9.6 }^{ +4.4 }_{ -2.8 }$  \\
\bottomrule
\end{tabular}
\label{tab:yields}
\end{table}

% End of text imported from the .//yield_table.tex input file

\begin{figure}[t]
\centering
\includegraphics[width=0.7\textwidth]{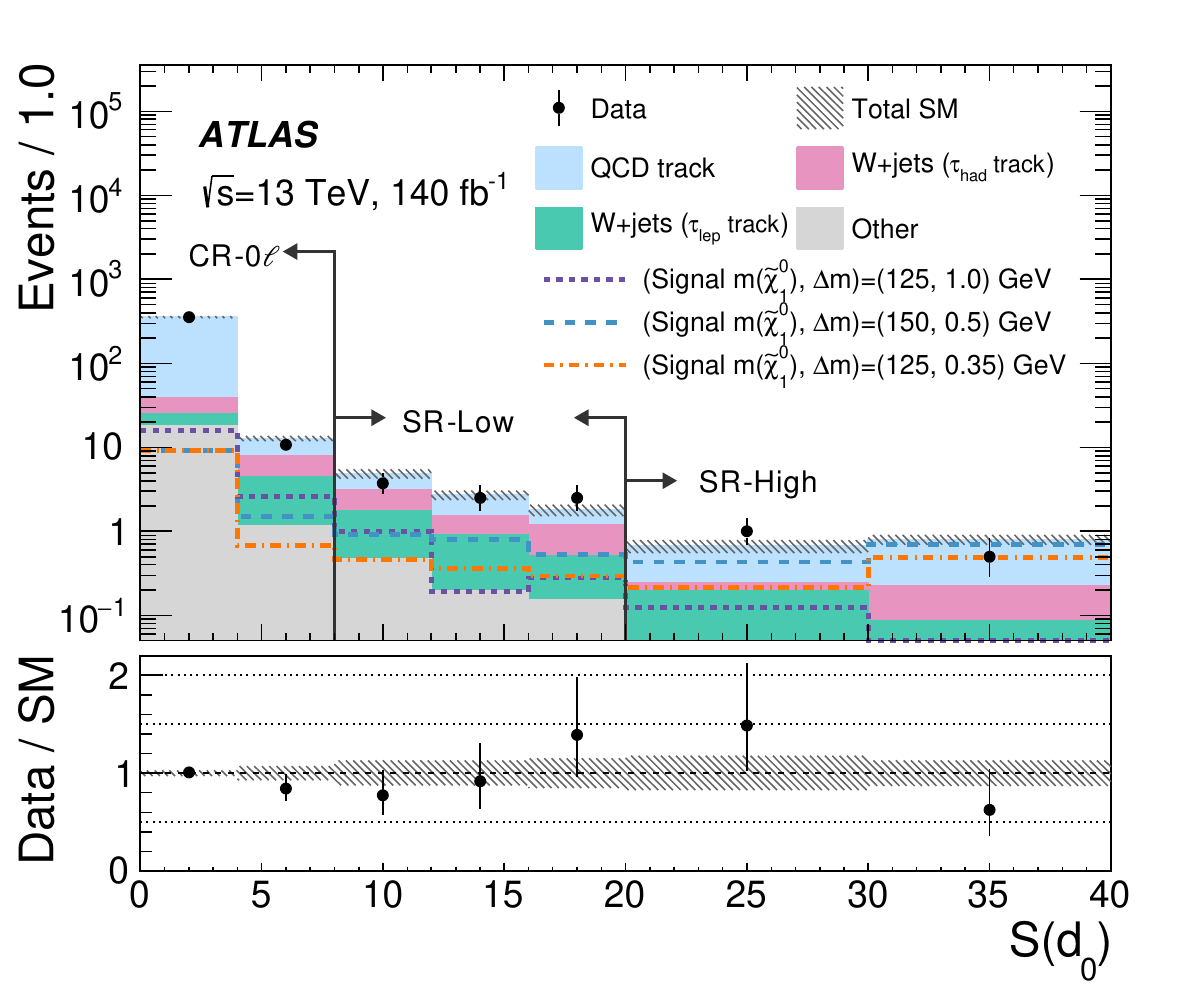}
\caption{Postfit \sdzero\ distribution in, and near, the SR and the adjacent control region CR-0$\ell$.
The selection criterion on the variable shown by each plot is removed, while the arrow indicates the cut value to define the region.
The SM background expectation is shown as a histogram stack,
and a few representative signal models are overlaid.
The last bin includes the overflow.
The hatched band indicates the uncertainty on the mean value of the SM background expectation, while the error bars on the black dots represent the Poisson error associated with the event yields.
}
\label{fig:SRplots}
\end{figure}
 
\FloatBarrier
 
\begin{figure}[ht]
\centering
\includegraphics[width=0.7\textwidth]{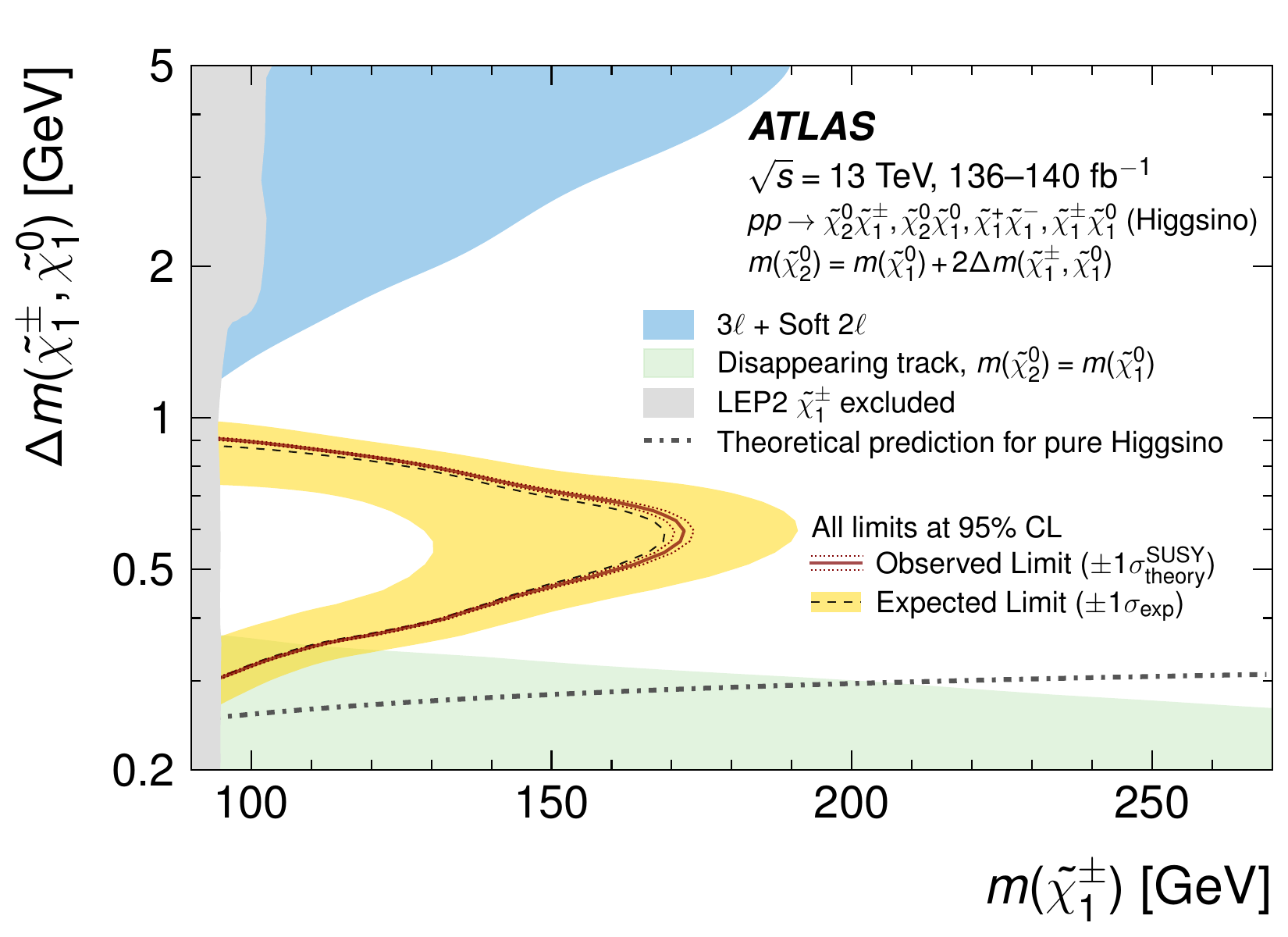}
\caption{
Expected (dashed black line) and observed (solid red line) 95\% CL exclusion limits on the Higgsino simplified model being considered.
These are shown with $\pm1\sigma_\text{exp}$ (yellow band) from experimental systematic and statistical uncertainties, and with $\pm1\sigma^{\text{SUSY}}_{\text{theory}}$ (red dotted lines) from signal cross section uncertainties, respectively.
The limits set by the latest ATLAS searches using the soft lepton ~\cite{SUSY-2018-16,SUSY-2019-09} and disappearing track~\cite{SUSY-2018-19} signatures are illustrated by the blue and green regions, respectively, while the limit imposed by the LEP experiments~\cite{LEPlimits} is shown in gray. The dot-dashed gray line indicates the predicted mass splitting for the pure Higgsino scenario~\cite{thomas1998phenomenology}.
}
\label{fig:limit}
\end{figure}

In conclusion, this Letter reports the results of a search for the pair production of nearly mass-degenerate
Higgsinos using 140~fb$^{-1}$ of $pp$ collision data at $\sqrt{s}=13~\TeV$ collected with the ATLAS detector at the LHC.
A novel signature is explored for the first time in LHC searches.
It features the use of a low-\pt displaced track to achieve sensitivity to a largely unconstrained region of the $0.3\text{--}1~\GeV$ Higgsino mass-splitting parameter range, which is challenging to probe with direct DM search experiments.
No excess above the SM expectation is observed and mass limits are set at 95\% CL within a simplified Higgsino model,
where Higgsino masses of up to about 170~\GeV are excluded, exceeding the limit set by the LEP experiments for the first time.
This result bridges a long-standing blind spot in the sensitivity of Higgsino searches, and establishes prospects for a conclusive test of the natural supersymmetry scenario, which predicts an electroweak-scale Higgsino mass.
 
\FloatBarrier
\section*{Acknowledgements}

% The next lines are included from the .//acknowledgements/Acknowledgements.tex input file

We thank CERN for the very successful operation of the LHC and its injectors, as well as the support staff at
CERN and at our institutions worldwide without whom ATLAS could not be operated efficiently.
 
The crucial computing support from all WLCG partners is acknowledged gratefully, in particular from CERN, the ATLAS Tier-1 facilities at TRIUMF/SFU (Canada), NDGF (Denmark, Norway, Sweden), CC-IN2P3 (France), KIT/GridKA (Germany), INFN-CNAF (Italy), NL-T1 (Netherlands), PIC (Spain), RAL (UK) and BNL (USA), the Tier-2 facilities worldwide and large non-WLCG resource providers. Major contributors of computing resources are listed in Ref.~\cite{ATL-SOFT-PUB-2023-001}.
 
We gratefully acknowledge the support of ANPCyT, Argentina; YerPhI, Armenia; ARC, Australia; BMWFW and FWF, Austria; ANAS, Azerbaijan; CNPq and FAPESP, Brazil; NSERC, NRC and CFI, Canada; CERN; ANID, Chile; CAS, MOST and NSFC, China; Minciencias, Colombia; MEYS CR, Czech Republic; DNRF and DNSRC, Denmark; IN2P3-CNRS and CEA-DRF/IRFU, France; SRNSFG, Georgia; BMBF, HGF and MPG, Germany; GSRI, Greece; RGC and Hong Kong SAR, China; ISF and Benoziyo Center, Israel; INFN, Italy; MEXT and JSPS, Japan; CNRST, Morocco; NWO, Netherlands; RCN, Norway; MEiN, Poland; FCT, Portugal; MNE/IFA, Romania; MESTD, Serbia; MSSR, Slovakia; ARRS and MIZ\v{S}, Slovenia; DSI/NRF, South Africa; MICINN, Spain; SRC and Wallenberg Foundation, Sweden; SERI, SNSF and Cantons of Bern and Geneva, Switzerland; MOST, Taipei; TENMAK, T\"urkiye; STFC, United Kingdom; DOE and NSF, United States of America.
 
Individual groups and members have received support from BCKDF, CANARIE, CRC and DRAC, Canada; CERN-CZ, PRIMUS 21/SCI/017 and UNCE SCI/013, Czech Republic; COST, ERC, ERDF, Horizon 2020, ICSC-NextGenerationEU and Marie Sk{\l}odowska-Curie Actions, European Union; Investissements d'Avenir Labex, Investissements d'Avenir Idex and ANR, France; DFG and AvH Foundation, Germany; Herakleitos, Thales and Aristeia programmes co-financed by EU-ESF and the Greek NSRF, Greece; BSF-NSF and MINERVA, Israel; Norwegian Financial Mechanism 2014-2021, Norway; NCN and NAWA, Poland; La Caixa Banking Foundation, CERCA Programme Generalitat de Catalunya and PROMETEO and GenT Programmes Generalitat Valenciana, Spain; G\"{o}ran Gustafssons Stiftelse, Sweden; The Royal Society and Leverhulme Trust, United Kingdom.
 
In addition, individual members wish to acknowledge support from CERN: European Organization for Nuclear Research (CERN PJAS); Chile: Agencia Nacional de Investigaci\'on y Desarrollo (FONDECYT 1190886, FONDECYT 1210400, FONDECYT 1230812, FONDECYT 1230987); China: Chinese Ministry of Science and Technology (MOST-2023YFA1605700), National Natural Science Foundation of China (NSFC - 12175119, NSFC 12275265, NSFC-12075060); Czech Republic: PRIMUS Research Programme (PRIMUS/21/SCI/017); EU: H2020 European Research Council (ERC - 101002463); European Union: European Research Council (ERC - 948254, ERC 101089007), Horizon 2020 Framework Programme (MUCCA - CHIST-ERA-19-XAI-00), European Union, Future Artificial Intelligence Research (FAIR-NextGenerationEU PE00000013), Italian Center for High Performance Computing, Big Data and Quantum Computing (ICSC, NextGenerationEU); France: Agence Nationale de la Recherche (ANR-20-CE31-0013, ANR-21-CE31-0013, ANR-21-CE31-0022), Investissements d'Avenir Labex (ANR-11-LABX-0012); Germany: Baden-Württemberg Stiftung (BW Stiftung-Postdoc Eliteprogramme), Deutsche Forschungsgemeinschaft (DFG - 469666862, DFG - CR 312/5-2); Italy: Istituto Nazionale di Fisica Nucleare (ICSC, NextGenerationEU); Japan: Japan Society for the Promotion of Science (JSPS KAKENHI JP21H05085, JSPS KAKENHI JP22H01227, JSPS KAKENHI JP22H04944, JSPS KAKENHI JP22KK0227); Netherlands: Netherlands Organisation for Scientific Research (NWO Veni 2020 - VI.Veni.202.179); Norway: Research Council of Norway (RCN-314472); Poland: Polish National Agency for Academic Exchange (PPN/PPO/2020/1/00002/U/00001), Polish National Science Centre (NCN 2021/42/E/ST2/00350, NCN OPUS nr 2022/47/B/ST2/03059, NCN UMO-2019/34/E/ST2/00393, UMO-2020/37/B/ST2/01043, UMO-2021/40/C/ST2/00187, UMO-2022/47/O/ST2/00148); Slovenia: Slovenian Research Agency (ARIS grant J1-3010); Spain: BBVA Foundation (LEO22-1-603), Generalitat Valenciana (Artemisa, FEDER, IDIFEDER/2018/048), Ministry of Science and Innovation (MCIN \& NextGenEU PCI2022-135018-2, MICIN \& FEDER PID2021-125273NB, RYC2019-028510-I, RYC2020-030254-I, RYC2021-031273-I, RYC2022-038164-I), PROMETEO and GenT Programmes Generalitat Valenciana (CIDEGENT/2019/023, CIDEGENT/2019/027); Sweden: Swedish Research Council (VR 2018-00482, VR 2022-03845, VR 2022-04683, VR grant 2021-03651), Knut and Alice Wallenberg Foundation (KAW 2017.0100, KAW 2018.0157, KAW 2018.0458, KAW 2019.0447, KAW 2022.0358); Switzerland: Swiss National Science Foundation (SNSF - PCEFP2\_194658); United Kingdom: Leverhulme Trust (Leverhulme Trust RPG-2020-004); United States of America: U.S. Department of Energy (ECA DE-AC02-76SF00515), Neubauer Family Foundation.

% End of text imported from the .//acknowledgements/Acknowledgements.tex input file

\appendix

% The next lines are included from the .//appendix.tex input file
 
\section*{Appendix A: Simulated data samples}
 
The signal samples are generated using the leading-order matrix elements with up to
two extra partons using \textsc{MadGraph}~\cite{Alwall:2014hca} v2.9.5
and interfaced with \textsc{Pythia}~8.306~\cite{Sjostrand:2014zea}
for simulating the subsequent decays, parton showering, and hadronization.
The A14 tune~\cite{ATL-PHYS-PUB-2014-021} of \textsc{Pythia} is used with the \texttt{NNPDF23lo} parton density function (PDF) set.
Matching between the matrix element and parton showering is performed following the CKKW-L prescription~\cite{Loennblad2012} with the merging scale set to $15~\GeV$.
The signal MC samples were processed with a fast simulation~\cite{SOFT-2010-01}, which relies on a parameterization of the calorimeter response~\cite{ATL-PHYS-PUB-2010-013}.
 
The $W+\text{jets}$ and $Z+\text{jets}$ backgrounds are generated using \textsc{Sherpa}~v2.2.11~\cite{Bothmann:2019yzt} with the \texttt{NNPDF30nnlo} PDF set~\cite{Ball:2014uwa}.
The diboson sample is generated with \textsc{Sherpa}~v2.2.2.
The \ttbar and single top-quark processes are generated at NLO with \textsc{Powheg-Box}~\cite{Frixione:2007nw,Nason:2004rx,Frixione:2007vw,Alioli:2010xd}~v2 and interfaced with \textsc{Pythia}~8.230 with the A14 tune.
The diagram removal scheme~\cite{Frixione:2008yi} is employed in the simulation of single top-quark production to account for interference between the \ttbar and $Wt$ processes.
The \texttt{NNPDF30nnlo} PDFs are used for all the MC samples of the SM processes.

\section*{Appendix B: Validation of background estimation}
 
This appendix describes the event selections used to validate the background estimates
introduced in the main text. See the Supplemental Material~\cite{SupplementalMaterial} associated with this Letter
for schematics to illustrate the region layout as well as tables summarizing the selection criteria.
 
\paragraph*{$\tau$ track background}
 
In the following, regions with a symbol $\tau_h$ or $\tau_\ell$ (referred to as $\tau_h$ regions or $\tau_\ell$ regions) are dominated by the hadronic and leptonic $\tau$ decays, respectively.
The $\tau_h$ ($\tau_\ell$) regions are defined by requiring no (one) lepton in the event.
In the $\tau_\ell$ regions, the electron (muon) is required to fail either of the following requirements; the tight identification criteria~\cite{EGAM-2018-01}, transverse impact parameter selection $|\dzero|/\sigma(\dzero) < 3$, or the ``Loose\_VarRad'' isolation criteria~\cite{atlas2023electron} (transverse impact parameter selection $|\dzero|/\sigma(\dzero) < 5$, $|\eta|<2.5$, or the ``PflowLoose'' isolation criteria~\cite{MUON-2018-03}).
This requirement enhances the contribution of leptons from $\tau$ decays with respect to the ordinary prompt leptons.
The \ptrecoil variable serves as a proxy for the \pt of the $W$ boson and refers to \met ($|\ptlepvec + \ptmissvec|$) in regions with no (one) lepton.
 
The $W(\ra \tau\nu)+\text{jets}$ MC sample is normalized to the data in CR-$\tau_h$ and CR-$\tau_\ell$ to obtain the estimate in the SR.
These CRs are defined by a higher track \pt requirement ($8\text{--}20~\GeV$) where the $\tau$ track background dominates with $>90~\%$ purity.
The \sdzero selection is loosened to acquire larger data statistics.
The extrapolation over the track \pt is validated in VR-$\tau_h$ and VR-$\tau_\ell$, where the intermediate track \pt range ($5\text{--}8~\GeV$) is selected.
To avoid signal contamination in the VRs, the \met selection is shifted to $300~\GeV<\met<400~\GeV$.
To validate the track \pt extrapolation in an isolated way, the estimates in the VRs are obtained by normalizing the MC in CRs with the same shifted \met requirement. These CRs are denoted CR2-$\tau_h$ and CR2-$\tau_\ell$.
The schematic of the region segmentation is shown in Supplemental Material~\cite{SupplementalMaterial}.

\paragraph{QCD track background}
 
Control region CR-1$\mu$, dominated by $W(\ra \mu\nu)+\text{jets}$, is used to extract the shape of the \sdzero distribution.
This is normalized to each CR ($\sdzero<8$) to get the estimate in the adjacent SR or VR ($\sdzero>8$).
Validation regions VR-$1e$, VR-$2\ell$ and VR-1$\gamma$ are defined with one electron, two leptons, and one photon, respectively,
where the corresponding dominant contributions are from $W(\ra e\nu)+\text{jets}$, $Z(\ra \ell\ell)+\text{jets}$, and $\gamma+\text{jets}$ production.
The choice of the VRs is motivated by their similar matrix element to that of the $Z(\ra \nu\nu)+\text{jets}$, the main physics process yielding the QCD track background in the SR.
The purities of $W/Z/\gamma+\text{jets}$ are $>90~\%$ except for VR(CR)-1$\gamma$ which has a $\sim30~\%$ contribution from QCD multijet production. However, the shape of the \sdzero distribution from these multijet events is found to be nearly identical to that of $\gamma+\text{jets}$ production.
The QCD track background considered in this fully data-driven estimate is therefore defined by events with the production of
$W(\ra \mu\nu)+\text{jets}$ in VR(CR)-1$\mu$;
$W(\ra e\nu)+\text{jets}$ in VR(CR)-1$e$;
$Z(\ra \ell\ell)+\text{jets}$ in VR(CR)-2$\ell$;
$\gamma+\text{jets}$ and multijet in VR(CR)-1$\gamma$.
The contribution from other backgrounds is modeled by the MC simulation in each VR and the associated CR, which is subtracted from the data in the \sdzero shape extraction or the normalization in the CRs.
The schematic of the region's segmentation are shown in Supplemental Material~\cite{SupplementalMaterial}.
 
The single-electron (single-photon) trigger is used to collect events with electrons (photons),
and the \met trigger is used to record events with large \met or muons.
Muons are counted as missing objects at the \met trigger.
Additional quality requirements are applied for electrons, muons and photons for regions requiring those;
the tight identification criteria~\cite{EGAM-2018-01}, transverse impact parameter selection $|\dzero|/\sigma(\dzero) < 3$ and Loose\_VarRad isolation criteria~\cite{atlas2023electron} are required for the electrons;
transverse impact parameter selection $|\dzero|/\sigma(\dzero) < 5$, $|\eta|<2.5$ and the PflowLoose isolation~\cite{MUON-2018-03} are imposed for the muons;
the tight isolation criteria~\cite{EGAM-2018-01} and $\pt>200~\GeV$ are applied for the photons.
These are to suppress the fake lepton and photon contributions from hadronic jets, or the nonprompt leptons from bottom and charm hadron decays.
Variable \ptrecoil is defined as the proxy for the \pt of the $W/Z/\gamma$ boson i.e.
$|\ptlepvec + \ptmissvec|$ in CR-1$\mu$ and VR(CR)-$1e$;
$|\ptleponevec + \ptleptwovec + \ptmissvec|$ in VR(CR)-$2\ell$;
$|\ptgammavec + \ptmissvec|$ in VR(CR)-1$\gamma$;
and \met in other regions.
The kinematic selection of the VRs is aligned as much as possible to the SR to ensure the same phase space is probed,
while a minimal set of additional selections are introduced to enhance the purity of $W/Z/\gamma+\text{jets}$ events in respective regions.

\paragraph{Hybrid validation region}
 
Validation region VR-0$\ell-$low\met is designed to validate the modeling of both the $\tau$ track and QCD track backgrounds, for which the contributions are nearly equal.
This VR is defined by shifting the \met selection to $300~\GeV<\met<400~\GeV$ with respect to the SR ($\met>600~\GeV$).
The $\tau$ track background is estimated by normalizing the MC in the low \met control regions (CR2-$\tau_h$ and CR2-$\tau_\ell$),
and the QCD track background is derived from the \sdzero shape that is simultaneously normalized in CR-0$\ell-$low\met.
 
\paragraph{Fit results}
 
Each VR is estimated by a separate profile likelihood fit,
except for the $\tau$ track VRs where the estimates of VR-$\tau_h$ and VR-$\tau_\ell$ are obtained from a single fit.
The associated CR(s) are used to constrain the likelihood,
including all the systematic uncertainties described in the main text, as well as the uncertainties on the lepton and photon isolation efficiencies.
Any potential signal contamination is expected to be below $5\%$ in all of the CRs and VRs and is therefore ignored in the fits.
The data and background estimates in the VRs are summarized in Figure~\ref{fig:pullVRs},
where they are shown to be consistent within the uncertainty.

\begin{figure}[thb!p]
\centering
\includegraphics[width=0.75\textwidth]{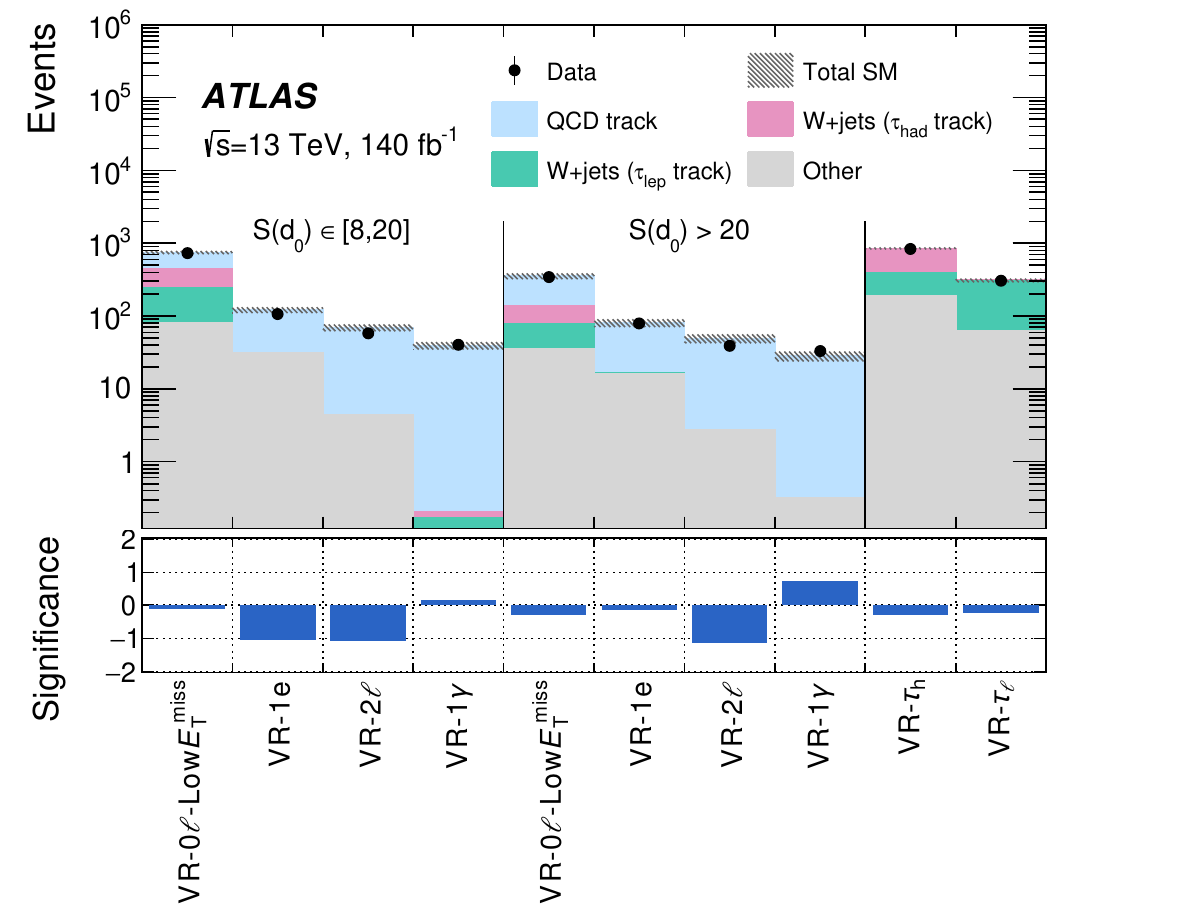}
\caption{Comparisons of the observed data and the expected SM background in the VRs after the profile likelihood fit in the CRs.
The hatched band indicates the total uncertainty on the SM background expectation, including both statistical and systematic uncertainties.
The bottom panel shows the statistical significances~\cite{Cousins:2007bmb} of the discrepancy between the expected and the observed event yields.
}
\label{fig:pullVRs}
\end{figure}
 
\section*{Appendix C: Postfit kinematic distributions in SR-High}
 
This appendix shows additional postfit distributions of kinematic variables important to the analysis. Figure~\ref{fig:N1plots_SRHigh} shows the \met and track \pt distributions in, and near, the phase space of SR-High.
 
\begin{figure}[thb!p]
\centering
\subfloat[]{\includegraphics[width=0.48\textwidth]{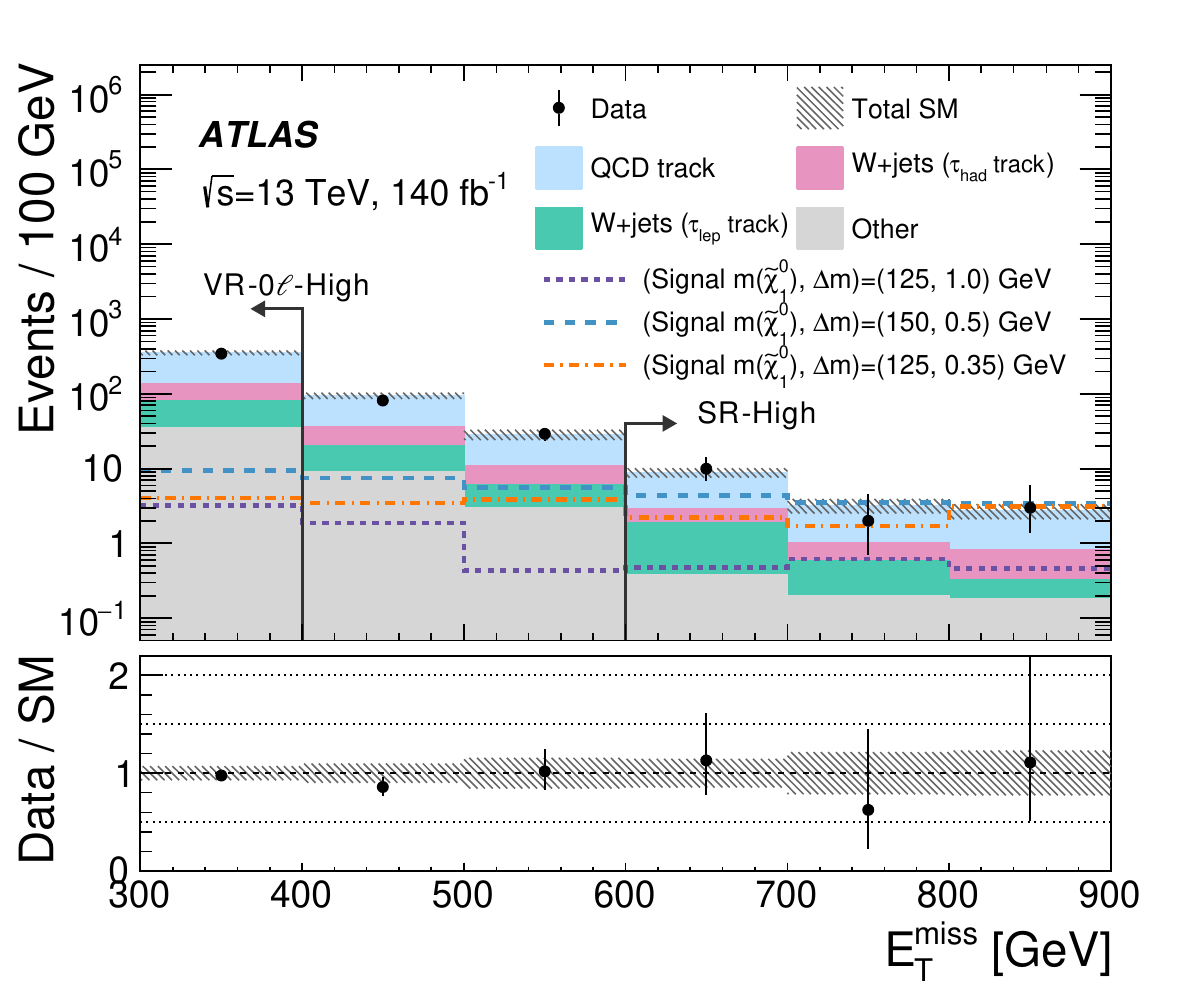}}
\subfloat[]{\includegraphics[width=0.48\textwidth]{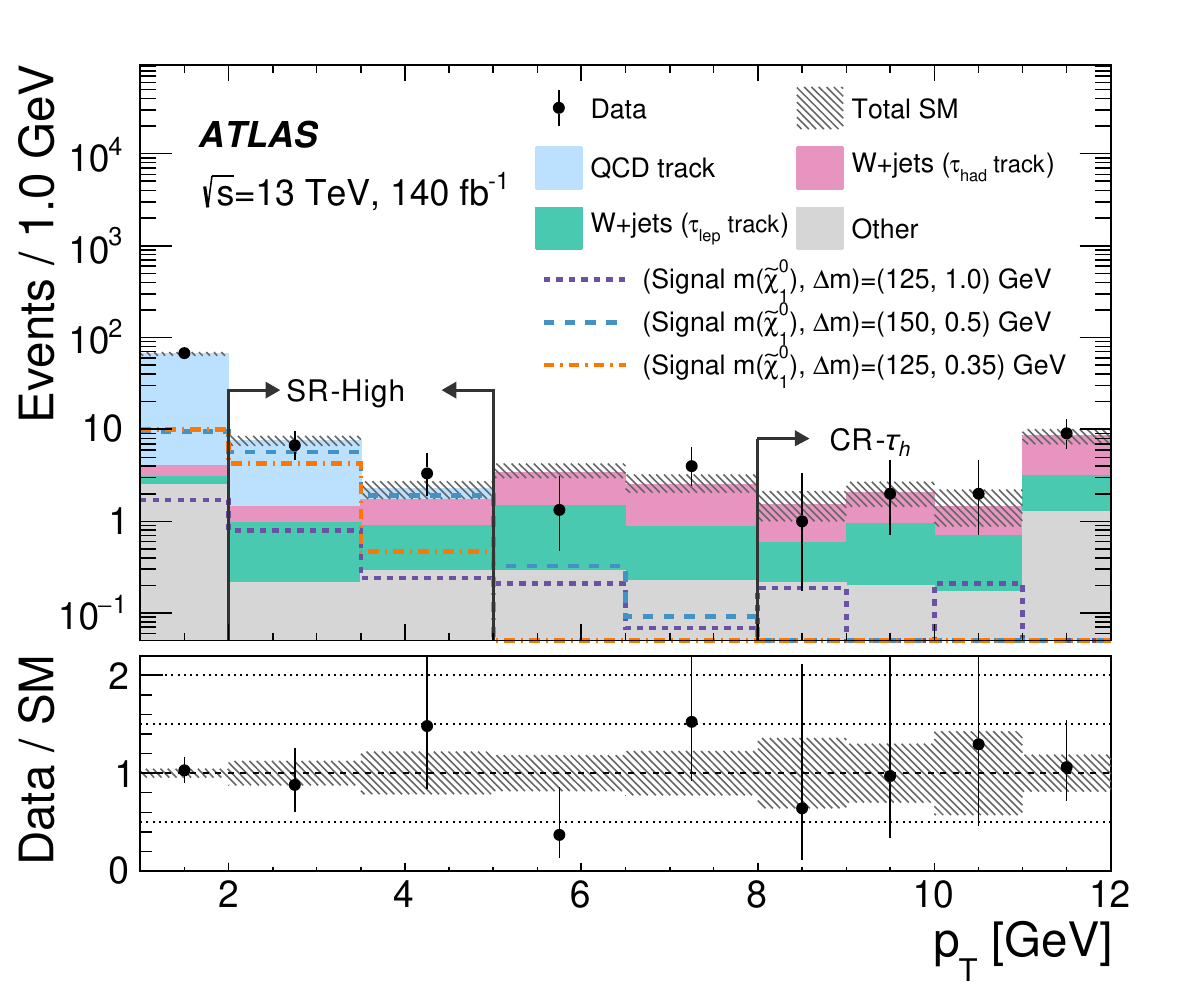}}
\caption{Distributions of \met and track \pt in, and near, the phase-space of SR-High.
The selection criterion on the variable shown by each plot is removed, while the arrow indicates the cut value to define the region.
The SM background expectation is shown as a histogram stack,
and a few representative signal models are overlaid. The last bin includes the overflow.
The hatched band indicates the uncertainty on the mean value of the SM background expectation, while the error bars on the black dots represent the Poisson error associated with the event yields.
}
\label{fig:N1plots_SRHigh}
\end{figure}

% End of text imported from the .//appendix.tex input file

\FloatBarrier
\printbibliography

\FloatBarrier
\part*{Supplemental Material}

% The next lines are included from the .//supplemental.tex input file

\begin{figure}[h]
\centering
\subfloat[]{\includegraphics[height=0.4\textwidth]{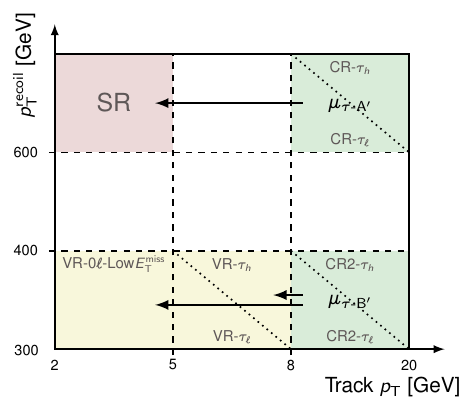} \label{fig:aux:TauRegions}}
\subfloat[]{\includegraphics[height=0.4\textwidth]{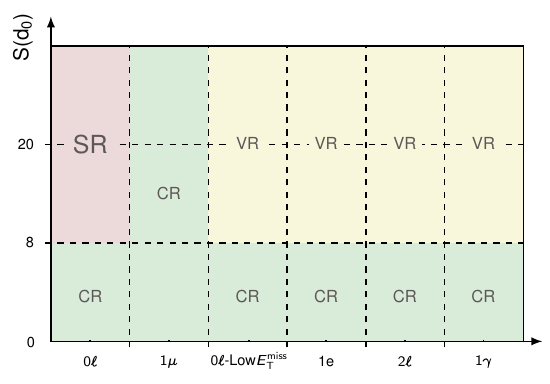} \label{fig:aux:QCDTrackRegions}}
\caption{Schematics illustrating the selection of the CRs and VRs used to estimate (a) the tau track background and (b) the QCD track background, along with the orthogonality with respect to the SR.
The arrows in (a) indicate the direction of extrapolation in the kinematic phase space which is validated in the VRs.
}
\end{figure}

\begin{table}[h]
\caption{Event selection for the control and validation regions relevant for estimating the tau track background, and that provide orthogonality with the SR.
Selections that are shared with SR are omitted.
Additional quality requirements are applied for leptons in the relevant regions, as described in the Appendix. Parentheses are used to group similar
control and validation regions together. The selections inside the parentheses apply to the control regions and indicate the criteria that enforces
orthogonality with respect to the corresponding validation region.
}
\centering
\begin{tabular}{lccccc} \toprule
Variable                           & SR                   & CR-$\tau_h$    & CR-$\tau_\ell$          & VR(CR2)-$\tau_h$  & VR(CR2)-$\tau_\ell$     \\
\midrule
$N_{\mathrm{\ell}}$                & = 0                  & = 0            & = 1                     & = 0               & = 1                     \\
$m_{\mathrm{T}}$ [GeV]             & --                   & --             & $<50$                   & --                & $<50$                   \\
\ptrecoil[GeV]                    & $>600$               & \multicolumn{2}{c}{$>600$}               & \multicolumn{2}{c}{[300,400]}               \\
Track \pt                          & [2,5]                & \multicolumn{2}{c}{[8,20]}               & \multicolumn{2}{c}{[5,8] ([8,20])}          \\
Track \sdzero                      & $>8$                 & \multicolumn{2}{c}{$>3$}                 & \multicolumn{2}{c}{$>3$}                    \\
\bottomrule
\end{tabular}
\end{table}

\begin{table}[h]
\caption{Event selection for the control and validation regions relevant for estimating the QCD track background, and that provide orthogonality with the SR.
Selections that are shared with SR are omitted.
The transverse mass ($m_{\mathrm{T}}$) is defined by $m_{\mathrm{T}} = \sqrt{2\pt(\ell)\met(1-\cos{\Delta \phi (\ell, \ptmissvec)})}$.
Additional quality requirements are applied for leptons and photons in the relevant regions, as described in the Appendix. Parentheses are used to group similar
regions together. The selections inside the parentheses apply to the control regions and indicate the criteria that enforces
orthogonality with respect to the corresponding signal or validation region.
}
\centering
\begin{adjustbox}{width=1.0\textwidth}
\begin{tabular}{lcccccc} \toprule
Variable                           & SR (CR-0$\ell$)      & CR-1$\mu$      & VR(CR)-0$\ell-$low\met  & VR(CR)-$1e$  & VR(CR)-$2\ell$       & VR(CR)-1$\gamma$ \\
\midrule
Trigger                            & \met                 & \met           & \met                    & Single-$e$   & \met or Single-$e$   & Single Photon  \\
$N(e)$                             & = 0                  & = 0            & = 0                     & = 1          & --                   & = 0            \\
$N(\mu)$                           & = 0                  & = 1            & = 0                     & = 0          & --                   & = 0            \\
$N(e~\text{or}~\mu)$               & = 0                  & = 1            & = 0                     & = 1          & = 2                  & = 0            \\
$N_{\mathrm{\gamma}}$              & = 0                  & = 0            & = 0                     & = 0          & = 0                  & = 1            \\
$\pt(\ell_1)$ [GeV]                & --                   & > 10           & --                      & > 30         & $\pt(\mu) > 10$ ($\pt(e) > 30$) & --  \\
$\pt(\ell_2)$ [GeV]                & --                   & --             & --                      & --           & > 10                 & -- \\
$m_{ll}$ [GeV]                     & --                   & --             & --                      & --           & [66.2, 116.2]        & -- \\
$m_{\mathrm{T}}$ [GeV]             & --                   & [56, 106]      & --                      & [56, 106]    & --                   & -- \\
\ptrecoil[GeV]                    & $>600$               & $>300$         & $[300,400]$             & $>300$       & $>300$               & $>600$ \\
Track \sdzero                      & $>8$ ($<8$)         & --             & \multicolumn{4}{c}{$>8$ ($<8$)}  \\
\bottomrule
\end{tabular}
\end{adjustbox}
\end{table}

% End of text imported from the .//supplemental.tex input file

\clearpage
% ATLAS Collaboration author list
% Reference date of SUSY-2020-04 is 2023-12-01
% Author list last updated on date 01-JUL-24
% Data extracted on 01-Jul-2024 for paper reference SUSY-2020-04
% at 8:23pm
 
\begin{flushleft}
\hypersetup{urlcolor=black}
{\Large The ATLAS Collaboration}

\bigskip

\AtlasOrcid[0000-0002-6665-4934]{G.~Aad}$^\textrm{\scriptsize 103}$,
\AtlasOrcid[0000-0001-7616-1554]{E.~Aakvaag}$^\textrm{\scriptsize 16}$,
\AtlasOrcid[0000-0002-5888-2734]{B.~Abbott}$^\textrm{\scriptsize 121}$,
\AtlasOrcid[0000-0002-1002-1652]{K.~Abeling}$^\textrm{\scriptsize 55}$,
\AtlasOrcid[0000-0001-5763-2760]{N.J.~Abicht}$^\textrm{\scriptsize 49}$,
\AtlasOrcid[0000-0002-8496-9294]{S.H.~Abidi}$^\textrm{\scriptsize 29}$,
\AtlasOrcid[0009-0003-6578-220X]{M.~Aboelela}$^\textrm{\scriptsize 44}$,
\AtlasOrcid[0000-0002-9987-2292]{A.~Aboulhorma}$^\textrm{\scriptsize 35e}$,
\AtlasOrcid[0000-0001-5329-6640]{H.~Abramowicz}$^\textrm{\scriptsize 152}$,
\AtlasOrcid[0000-0002-1599-2896]{H.~Abreu}$^\textrm{\scriptsize 151}$
% ...
\bigskip
\\

% ...
$^{4}$LAPP, Université Savoie Mont Blanc, CNRS/IN2P3, Annecy; France.\\
% ...
$^{16}$Department for Physics and Technology, University of Bergen, Bergen; Norway.\\
% ...
$^{29}$Physics Department, Brookhaven National Laboratory, Upton NY; United States of America.\\
% ...
$^{35}$$^{(a)}$Facult\'e des Sciences Ain Chock, R\'eseau Universitaire de Physique des Hautes Energies - Universit\'e Hassan II, Casablanca;$^{(b)}$Facult\'{e} des Sciences, Universit\'{e} Ibn-Tofail, K\'{e}nitra;$^{(c)}$Facult\'e des Sciences Semlalia, Universit\'e Cadi Ayyad, LPHEA-Marrakech;$^{(d)}$LPMR, Facult\'e des Sciences, Universit\'e Mohamed Premier, Oujda;$^{(e)}$Facult\'e des sciences, Universit\'e Mohammed V, Rabat;$^{(f)}$Institute of Applied Physics, Mohammed VI Polytechnic University, Ben Guerir; Morocco.\\
% ...
$^{44}$Physics Department, Southern Methodist University, Dallas TX; United States of America.\\
% ...
$^{49}$Fakult\"{a}t Physik , Technische Universit{\"a}t Dortmund, Dortmund; Germany.\\
% ...
$^{55}$II. Physikalisches Institut, Georg-August-Universit\"{a}t G\"ottingen, G\"ottingen; Germany.\\
% ...
$^{103}$CPPM, Aix-Marseille Universit\'e, CNRS/IN2P3, Marseille; France.\\
% ...
$^{121}$Homer L. Dodge Department of Physics and Astronomy, University of Oklahoma, Norman OK; United States of America.\\
% ...
$^{151}$Department of Physics, Technion, Israel Institute of Technology, Haifa; Israel.\\
$^{152}$Raymond and Beverly Sackler School of Physics and Astronomy, Tel Aviv University, Tel Aviv; Israel.\\
% ...
$^{a}$ Also Affiliated with an institute covered by a cooperation agreement with CERN.\\
$^{b}$ Also at An-Najah National University, Nablus; Palestine.\\
$^{c}$ Also at Borough of Manhattan Community College, City University of New York, New York NY; United States of America.\\
$^{d}$ Also at Center for High Energy Physics, Peking University; China.\\
$^{e}$ Also at Center for Interdisciplinary Research and Innovation (CIRI-AUTH), Thessaloniki; Greece.\\
$^{f}$ Also at Centro Studi e Ricerche Enrico Fermi; Italy.\\
% ...

\end{flushleft}

% Created with Glance <Atlas.Glance@cern.ch>

\end{document}